\documentclass[12pt]{article}
\usepackage[top=25truemm,bottom=27truemm,left=22truemm,right=22truemm]{geometry}

\usepackage{indentfirst}
\usepackage[dvipdfmx]{graphicx}
\usepackage{float}
\usepackage{url}
\usepackage{siunitx}
\usepackage{array}
\usepackage{amsmath}
\usepackage{amssymb}
\usepackage{fancybox}
\usepackage{ascmac}
\usepackage{bm}
\usepackage{mathrsfs}
\usepackage{cancel}
\usepackage{physics}
\usepackage{color}
\usepackage{authblk}
\usepackage{appendix}
\usepackage{comment}
\usepackage{mhchem}
\usepackage{caption,latexsym,amsfonts,mathtools,here}
\usepackage{subcaption}
\usepackage{hyperref}
\usepackage{multirow}

\numberwithin{equation}{section}

\newcommand{\beqn}{\begin{eqnarray}}
\newcommand{\eeqn}{\end{eqnarray}}

\begin{document}
\begin{titlepage}
\begin{flushright}
{YITP-26-92, KOBE-COSMO-26-07}
\end{flushright}

\vspace{50pt}

\begin{center}

{\large{\textbf{Intermittency in Quantum Graviton-Phonon Conversion}}}

\vspace{25pt}

{Yuna Gouin$^{\flat\dagger}$, Sugumi Kanno$^{*\natural}$, Jiro Soda$^{\flat}$}
\end{center}

\vspace{20pt}

\shortstack[l]
{\hspace{1.6cm}\it {\small $^\dagger$Département de Physique, Ecole Normale Supérieure - PSL, Paris 75005, France} \\[5pt]
\it \hspace{1.6cm}$^\flat$ {\small Department of Physics, Kobe University, Kobe 657-8501, Japan} \\[5pt]
\hspace{1.6cm}\it {\small $^*$Department of Physics, Kyushu University, Fukuoka 819-0395, Japan} \\[5pt]
\hspace{1.6cm}\it {\small $^\ddag$ Quantum and Spacetime Research Institute, Kyushu University}\\[5pt]
\hspace{1.6cm}\it {\small $^\natural$ Center for Gravitational Physics and Quantum Information,} \\[2pt]
\hspace{1.9cm}\it {\small Yukawa Institute for Theoretical Physics, Kyoto University,} \\[5pt]
}
\vspace{28pt}

\begin{abstract}
A graviton can be converted into a phonon in a resonant bar detector. First-order perturbation theory predicts a strong enhancement of this conversion for coherent and squeezed graviton states, but the probability can exceed unity when the coherent
or squeezing parameter is large. Since a conversion probability must satisfy the unitarity bound, we solve the graviton-phonon quantum dynamics exactly within the rotating-wave approximation.
 For an initial coherent state, we find that the conversion occurs intermittently through narrow bursts separated by intervals of strong suppression. For an initial squeezed state, the departure from perturbative behavior occurs earlier, and the conversion is strongly suppressed after its initial growth. These effects may provide signatures of quantum graviton-phonon dynamics relevant to single-graviton detection.
\end{abstract}
\end{titlepage}

\tableofcontents

\section{Introduction}

In 2015, LIGO made the first direct detection of gravitational waves~\cite{LIGOScientific:2016aoc}. Since then, gravitational-wave astronomy and multimessenger astronomy have developed rapidly~\cite{Arimoto:2021cwc}. 
Given these developments, it is natural to search for signatures of gravitons from the perspective of quantum gravity. Primordial gravitational waves generated during inflation are generally believed to have a quantum origin.
One possible approach to searching for gravitons is therefore to
investigate the quantum properties of primordial gravitational waves~\cite{Grishchuk:1989ss,Grishchuk:1990bj,Albrecht:1992kf}.
Indeed, several previous studies have explored their quantum nature using measures such as entanglement entropy~\cite{Kanno:2014lma}, entanglement negativity~\cite{Kanno:2014bma}, entanglement discord~\cite{Kanno:2016gas}, violations of Bell inequalities~\cite{Maldacena:2015bha,Kanno:2017dci}, and second-order coherence~\cite{Kanno:2018cuk, Kanno:2019gqw}. 
However, designing a concrete experimental setup remains challenging. 

Dyson argued that direct detection of a single graviton is practically extremely difficult~\cite{Dyson:2013hbl}. 
Two main research directions have been pursued to overcome this difficulty. Since the challenge is related to the large occupation number of gravitons at experimentally accessible frequencies, one approach is to extend gravitational-wave observations into the high-frequency regime. Searches for high-frequency gravitational waves have therefore been initiated~\cite{Ito:2019wcb,Ito:2020wxi}.
Another approach is to detect gravitons indirectly through graviton-induced noise~
\cite{Parikh:2020nrd,Kanno:2020usf,Parikh:2020kfh,Parikh:2020fhy}.
For example, the decoherence of spatially superposed mirrors caused by graviton noise may provide a means of probing gravitons~\cite{Kanno:2021gpt}. 
To reveal their quantum nature, it may also be useful to focus on the squeezed quantum states of gravitational radiation~
\cite{Guerreiro:2025sge,Manikandan:2025ykr,Manikandan:2025qgv,Manikandan:2025hlz}.
The generation of squeezed gravitational waves by superradiant axion fields has been studied in~ \cite{Dorlis:2025zzz,Dorlis:2025amf}. 
Deviations from a coherent-state description of gravitational waves emitted by binary black holes have also been estimated~\cite{Kanno:2025how,Das:2025kyn}, and  the squeezing parameter has been evaluated for quasinormal modes~\cite{Manikandan:2025dea}.
It has been pointed out that observations of the quantum properties of gravitational waves from binary black holes may contain information about the early universe~\cite{Kanno:2025fpz}.

In this paper, we focus on the proposal by Tobar et al.~\cite{Tobar:2023ksi} to detect single gravitons using gravitational-wave bar detectors~\cite{Preparata:1988am,Grishchuk:1992gn}.
A key ingredient of this proposal is the coherent-state description of gravitational waves. 
It has been shown that the graviton-to-phonon conversion probability can be enhanced by the squared magnitude of the coherent-state parameter. However, this result was obtained using first-order perturbation theory.
For this reason, such an enhancement has been argued not to constitute unambiguous evidence for quantization of gravity~\cite{Carney:2023nzz,Carney:2024dsj}.
To establish the quantum nature of gravitons more directly, the graviton-number statistics have therefore been investigated in detail~\cite{Toccacelo:2026hcz}.
Our aim is similar, but our approach is different: we probe the quantum nature of gravitons by reexamining the graviton-phonon conversion process beyond first-order perturbation theory. 

In this paper, we investigate the detection of single gravitons within a field-theoretical framework~\cite{Ikeda:2025qac,Ikeda:2025uae,Arani:2026jyz}. Previous analyses have relied on either a semiclassical approximation or first-order perturbation theory. Here, instead, we take the full quantum dynamics into account. We first demonstrate that the conversion probability satisfies the unitarity bound. We then identify a distinctive feature of its time evolution, namely, intermittent conversion. This behavior may provide a signature of the underlying quantum dynamics through single-graviton detection.

The remainder of this paper is organized as follows.
In section~2, we explain how to solve the Schrödinger equation by reducing the quantum dynamics to a set of ordinary differential equations.
In section~3, we calculate the conversion probabilities for single-graviton, coherent, and squeezed initial states. We show that the conversion probability always satisfies the unitarity bound. We also find that intermittent behavior emerges for a large coherent-state parameter, while strong suppression occurs for the squeezed state.
Section~4 is devoted to our conclusions.

\section{Exact analysis of quantum dynamics}

We begin with the Schrödinger equation
\begin{align}
    i\frac{d}{d t} \ket{\psi(t)} = H(t)\ket{\psi(t)} \ .
\end{align}
The formal solution is given by
\begin{align}
    \ket{\psi (t)}= U(t)\ket{\psi(0)}
    \  , \qquad U(t) = \mathcal{T} \exp \left[ -i \int_{0}^t H(t') dt'\right] \ ,
\end{align}
where $\mathcal{T}$ denotes the time-ordering operator. In general, the time-ordered exponential encodes the full quantum dynamics, but
extracting explicit physical information from it is not straightforward. A standard approach is to employ perturbation theory. However, the first-order result coincides with the corresponding semiclassical result and therefore does not, by itself, establish that single-graviton detection
requires a quantum description of gravity.
Here, we instead employ an algebraic method that allows us to describe the full quantum dynamics.
This approach reduces the problem to a set of
ordinary differential equations which can then be solved numerically.

\subsection{Time evolution operator}

Let us consider the exact dynamics of a coupled graviton-phonon system. We assume that the wavelength of gravitational waves is much larger than the size of the detector. Under this assumption, the gravitational
wave can be approximated by a single mode. The Hamiltonian is then given
by~\cite{Tobar:2023ksi}:
\begin{align}
    H = \omega_a a^\dagger a  + \omega_b b^\dagger b  - g (a + a^\dagger)( b+ b^\dagger  ) \ ,
\end{align}
where the operators $a$ and $a^\dagger$
are the annihilation and creation operators for the graviton mode, while
$b$ and $b^\dagger$ are those for the phonon mode. They satisfy 
\begin{align}
    [a,a^\dagger]=1,
    \qquad
    [b,b^\dagger]=1,
\end{align}
with all mixed commutators vanishing.
Here, the coupling constant is
\begin{align}
    g =   \sqrt{\frac{8\pi G M\omega_a^3 L^2}{\omega_b \pi^4 V}} \ ,
\end{align}
$G$ is Newton's constant, $V$ is a characteristic volume associated with the gravitational-wave mode, and $M$ and $L$ denote the mass and length of the cylindrical bar, respectively. For example, for a detector with mass $M=10^3\,\mathrm{kg}$, length
$L=10\,\mathrm{m}$, frequencies $\omega_a=\omega_b=1\,\mathrm{Hz}$,
and $V\simeq\omega_a^{-3}$, we obtain
$g\simeq10^{-17}\,\mathrm{Hz}$. We take these values as our benchmark
parameters.

The numerical parameters used in
Figs.~\ref{fig:coherent-comparison}--\ref{result}, however, differ from
these benchmark values and are chosen for illustrative purposes. They are
selected to make the breakdown of the perturbative approximation, the
intermittent behavior, and the difference between coherent and squeezed
initial states clearly visible. Accordingly, they should not be interpreted
as representing a specific realistic detector configuration.

For sufficiently weak coupling, we adopt the rotating-wave approximation.
In the interaction picture, the counter-rotating terms $ab$ and
$a^\dagger b^\dagger$ oscillate with frequency $\omega_a+\omega_b$.
Near resonance and provided that $g\ll \omega_a,\omega_b$, their
contributions average to zero over the relevant interaction timescale.
The Hamiltonian therefore reduces to
\begin{align}
    H = \omega_a a^\dagger a  + \omega_b b^\dagger b  
    - g (a b^\dagger + a^\dagger b ) \ .
\end{align}
The operators appearing in this Hamiltonian form a closed Lie algebra. Therefore, the time-evolution operator can be factorized into a product of
unitary operators~\cite{Schumaker:1986tlu}:
\begin{align}
    U(t)= e^{i\delta (t) }R_aR_b T\,,
\end{align}
where $R_a$ and $R_b$ are single-mode rotation operators, and $T$ is a
two-mode mixing operator:
\begin{align}
    &R_a=\exp(-i\theta_a(t) a^\dagger a),\qquad 
    R_b=\exp(-i\theta_b (t) b^\dagger b), \nonumber \\
    & T=\exp\left[ q(t) (e^{-i\chi(t)}ab^\dagger -e^{i\chi(t)}a^\dagger b)\right] \ . 
\end{align}
Using the Hadamard lemma,
\begin{align}
    e^ABe^{-A}= B + [A,B] + \frac{1}{2!}[A,[A,B]]+ \dots \ ,
\end{align}
we obtain the following transformation rules:
\begin{align} \label{commutators}
    R_aaR_a^\dagger =& e^{i\theta_a}a \ ,\qquad &R_aa^\dagger R_a^\dagger =& e^{-i\theta_a}a^\dagger \ ,\nonumber \\ 
    R_bbR_b^\dagger =& e^{i\theta_b}b \ ,\qquad &R_bb^\dagger R_b^\dagger =& e^{-i\theta_b}b^\dagger \ ,\nonumber \\
    TaT^\dagger =& \cos q\, a+e^{i\chi}\sin q\,b \ ,\qquad 
    &Ta^\dagger T^\dagger =& \cos q\, a^\dagger+e^{-i\chi}\sin q\,b^\dagger \ ,\nonumber \\
    TbT^\dagger =& \cos q\, b-e^{-i\chi}\sin q\,a \ ,\qquad 
    &Tb^\dagger T^\dagger =& \cos q\, b^\dagger-e^{i\chi}\sin q\,a^\dagger \ .
\end{align}

The time-evolution operator satisfies the Schrödinger equation
\begin{align} \label{Schrodinger}
    i\dot{U}U^\dagger=H      \ ,
\end{align}
where a dot denotes differentiation with respect to time.
Using the
factorized form of $U(t)$, the left-hand side of Eq.~\eqref{Schrodinger} becomes
\begin{eqnarray}
    i\dot{U}U^\dagger = -\dot{\delta} + i\dot{R_a}R_a^\dagger + i\dot{R_b}R_b^\dagger  
    + R_aR_b(i\dot{T}T^\dagger ) R_b^\dagger R_a^\dagger  \ .
\end{eqnarray}
Each term can be evaluated using
Eq.~(\ref{commutators}) and the identity
\begin{align}
    \partial_t(e^{f(t)})e^{-f(t)} = \dot{f} + \frac{1}{2!}[f,\dot{f}]+ \frac{1}{3!}[f,[f,\dot{f}]]+ \cdots  \ .
\end{align}
We then obtain
\begin{eqnarray}
  i\dot{R_a}R_a^\dagger &=& \dot{\theta}_aa^\dagger a\ , \nonumber \\
  \qquad i\dot{R_b}R_b^\dagger  &=& \dot{\theta}_b b^\dagger b \ , \nonumber\\
  R_aR_b(i\dot{T}T^\dagger ) R_b^\dagger R_a^\dagger
 &=& \left(i\dot{q} + \frac{1}{2}\dot{\chi} \sin 2q \right) e^{-i\chi} e^{i(\theta_a -\theta_b)} a b^\dagger 
 + \left(-i\dot{q} + \frac{1}{2}\dot{\chi} \sin 2q \right) 
 e^{i\chi} e^{-i(\theta_a -\theta_b)}  a^\dagger  b     \nonumber\\
&& +\frac{1}{2}\dot{\chi} (\cos 2q -1) (a^\dagger a - b^\dagger b)\ .
\end{eqnarray}
Matching the coefficients of the independent operators on both sides of Eq. \ref{Schrodinger}, we obtain the following five coupled equations:
\begin{eqnarray}
    \dot{\delta}&=&0  \ ,\\
    \dot{\theta_a}&=& \omega_a -\frac{1}{2}\dot{\chi}\left(\cos 2q -1\right) \ ,\\
     \dot{\theta_b}&=& \omega_b +\frac{1}{2}\dot{\chi}\left(\cos 2q -1\right) \ ,\\
     \dot{q} &=& -g \sin\chi \cos (\theta_a -\theta_b) 
     + g \cos \chi \sin (\theta_a -\theta_b) \ ,\\
     \dot{\chi} \sin 2q &=& 
     -2g \cos\chi \cos (\theta_a - \theta_b ) 
     -2g \sin \chi \sin (\theta_a -\theta_b)  \ .
\end{eqnarray}
Thus, the full quantum dynamics is reduced to a system of coupled ordinary differential equations.

\subsection{Time evolution equations for the parameters}

We now solve the equations governing the time evolution of the parameters.
The equation for $\delta$ is trivial, and we set  $\delta=0$ henceforth. Introducing the combinations
\begin{eqnarray}
  \bar{\theta}=  \theta_a + \theta_b\ , 
  \qquad \theta = \theta_a -\theta_b \ ,
\end{eqnarray}
we immediately find
\begin{eqnarray}
  \bar{\theta}=  \left(\omega_a + \omega_b\right) t \,,
\end{eqnarray}
where we have chosen $\bar{\theta}(0)=0$. 

The remaining equations exhibit a coordinate singularity at $q=0$. To avoid this singularity, it is convenient to introduce the Cartesian-like variables
\begin{eqnarray}
  x= \sin 2q \cos \chi \ , \qquad y = \sin 2q \sin \chi \ .
\end{eqnarray}
In terms of these variables, the remaining equations form the following closed system:
\begin{eqnarray}
    \dot{\theta}&=& \omega_a -\omega_b 
    -2g \cos \theta \frac{x}{1+\sqrt{1-x^2-y^2}}
    -2g \sin \theta \frac{y}{1+\sqrt{1-x^2-y^2}}
    \ , \\
     \dot{x} &=& 
    2g \cos \theta \frac{xy}{1+\sqrt{1-x^2-y^2}}
    +2g \sin \theta \left[1-\frac{x^2}{1+\sqrt{1-x^2-y^2}} \right] \ ,\\
     \dot{y} &=& 
    -2g \sin \theta \frac{xy}{1+\sqrt{1-x^2-y^2}}
    -2g \cos \theta \left[1-\frac{y^2}{1+\sqrt{1-x^2-y^2}} \right] \ .
\end{eqnarray}
In deriving these equations, we have restricted the parameter range to
\begin{equation}
    0\leq q\leq \frac{\pi}{4}.
\end{equation}
We have verified that this condition is satisfied throughout our numerical calculations.

The above equations can be solved numerically. Nevertheless, it is instructive to examine their perturbative solutions. At zeroth order in
the coupling constant $g$, we obtain
\begin{eqnarray}
  \theta^{(0)} = ( \omega_a -\omega_b) t \equiv \Delta t  \ , \quad
  x^{(0)} =0 \ ,\quad 
  y^{(0)}=0   \,,
\end{eqnarray}
where $\Delta\equiv\omega_a-\omega_b$ denotes the detuning.
At first order in $g$, the solutions are
\begin{eqnarray}
  \theta^{(1)} = 0 \ , \quad
  x^{(1)} =  - \frac{2g}{\Delta} \left(1-\cos\Delta t\right)\ ,\quad 
  y^{(1)}=- \frac{2g}{\Delta} \sin \Delta t  \,,
\end{eqnarray}
where we have imposed the initial conditions
$x(0)=y(0)=0$.
In the resonant limit, $\Delta\to0$, these expressions reduce to
\begin{align}
    \theta^{(1)}(t) &=0\,,
    &
    x^{(1)}(t) &= 0\,,
    &
    y^{(1)}(t) &= -2gt \,.
\end{align}
Using $x=\sin 2q\cos\chi$ and
$y=\sin 2q\sin\chi$, we therefore obtain, to leading order in $g$,
\begin{eqnarray}
  \theta(t)  = \Delta t \ , \quad
  q(t)  = gt \ ,\quad 
  \chi(t) = - \frac{\pi}{2}  \ .\label{1st}
\end{eqnarray}

\section{Conversion probability, unitarity bound, and intermittency}
As discussed in Sec.~2, we work within the rotating-wave approximation and neglect the counter-rotating terms $ab$ and $a^\dagger b^\dagger$. We now analyze the graviton-phonon conversion probability for several initial quantum states.

At the most elementary level, the conversion process corresponds to a single graviton being converted into a single phonon. In realistic observations, however, gravitational waves are typically described as
classical fields. Within quantum theory, a classical gravitational wave can be represented by a coherent state. It is therefore natural to consider
the graviton coherent state
\begin{eqnarray}
\ket{\alpha}=D(\alpha)\ket{0} \ , \qquad
D(\alpha) =e^{\alpha a^\dagger-\alpha^*a}
\end{eqnarray}
as an initial state.
We also consider the squeezed-vacuum state
\begin{eqnarray}
\ket{\xi}=S(\xi)\ket{0} \ , \qquad
S(\xi) =e^{\frac{1}{2} \xi a^2- \frac{1}{2} \xi^* a^{\dagger 2}}
\ , \qquad \xi = r e^{-i\phi} \ .
\end{eqnarray}
In the following subsections, we compute the corresponding transition amplitudes for single-graviton, coherent, and squeezed initial states.

\subsection{Single-graviton state}

Consider an initial state containing a single graviton and no phonons.
The amplitude for converting the graviton into a single phonon is
\begin{align}
    \mathcal{A}_{g\to p}(t)
    &\equiv
    \bra{0}b\,U(t)\,a^\dagger\ket{0}
    \nonumber\\
    &=
    \bra{0}b\,R_aR_bT\,a^\dagger\ket{0}
    \nonumber\\
    &=
    e^{-i\theta_b}
    \bra{0}b
    \left(
        \cos q\,a^\dagger
        +e^{-i\chi}\sin q\,b^\dagger
    \right)
    T\ket{0}
    \nonumber\\
    &=
    e^{-i(\theta_b+\chi)}\sin q .
\end{align}
Here, we have used
\begin{align}
bR_b=e^{-i\theta_b} R_b b\,,\qquad T\ket{0}=\ket{0}\,,\qquad \bra{0}R_a=\bra{0}R_b=\bra{0}\,.
\end{align}
The corresponding conversion probability is therefore
\begin{align}
    P_{g\to p}(t)
    =
    \left|
        \bra{0}b\,U(t)\,a^\dagger\ket{0}
    \right|^2
    =
    \sin^2 q(t).
\end{align}
Using the leading-order resonant solution $q(t)=gt$, we recover
\begin{align}
    P_{g\to p}(t)=\sin^2(gt).
    \label{leading}
\end{align}
This oscillatory behavior agrees with the result obtained from the classical analysis of two coupled modes.

\subsection{Coherent state}

We next consider the initial two-mode state
\begin{align}
    \ket{\alpha,0}
    \equiv
    \ket{\alpha}_a\otimes\ket{0}_b,
\end{align}
where the graviton mode is in a coherent state and the phonon mode is in
the vacuum. Under free evolution, the coherent-state parameter evolves as
\begin{align}
    \alpha(t)=\alpha e^{-i\omega_a t}.
\end{align}
We calculate the transition amplitude from $\ket{\alpha,0}$ to
$\ket{\alpha(t),1}$:
\begin{align}
    \mathcal{A}_{\alpha\to\alpha,1}(t)
    &\equiv
    \bra{\alpha(t),1}U(t)\ket{\alpha,0}
    \nonumber\\
    &=
    \bra{\alpha(t),0}b\,U(t)\ket{\alpha,0}
    \nonumber\\
    &=
    \bra{\alpha e^{-i\omega_a t},0}
    b\,R_aR_bT
    \ket{\alpha,0}
    \nonumber\\
    &=
    e^{-i\theta_b}
    \bra{\alpha e^{-i\omega_a t}e^{i\theta_a},0}
    T
    \left(
        \cos q\,b
        +e^{-i\chi}\sin q\,a
    \right)
    \ket{\alpha,0}
    \nonumber\\
    &=
    e^{-i(\theta_b+\chi)}
    \alpha\sin q\,
    \bra{\alpha e^{-i\omega_a t}e^{i\theta_a},0}
    T
    \ket{\alpha,0}
    \nonumber\\
    &=
    e^{-i(\theta_b+\chi)}
    \alpha\sin q\,
    \bra{\alpha e^{-i\omega_a t}e^{i\theta_a},0}
    T D_a(\alpha)T^\dagger
    \ket{0,0}
    \nonumber\\
    &=
    e^{-i(\theta_b+\chi)}
    \alpha\sin q\,
    \left\langle
        \alpha e^{-i\omega_a t}e^{i\theta_a}
        \middle|
        \alpha\cos q
    \right\rangle_a
    \left\langle
        0
        \middle|
        \alpha e^{-i\chi}\sin q
    \right\rangle_b .
\end{align}
Here, we have used
\begin{align}
    T D_a(\alpha)T^\dagger
    =
    D_a(\alpha\cos q)
    D_b(\alpha e^{-i\chi}\sin q).
\end{align}
Using the coherent-state overlap
\begin{align}
    \left|\langle\alpha|\beta\rangle\right|^2
    =
    e^{-|\alpha-\beta|^2},
\end{align}
we obtain
\begin{align}
    P_{\alpha\to\alpha,1}(t)
    &\equiv
    \left|
        \bra{\alpha(t),1}U(t)\ket{\alpha,0}
    \right|^2
    \nonumber\\
    &=
    \left|
        \bra{\alpha(t),0}b\,U(t)\ket{\alpha,0}
    \right|^2
    \nonumber\\
    &=
    |\alpha|^2\sin^2 q(t)
    \exp\left[
        -2|\alpha|^2
        \left[
            1-\cos q(t)
            \cos\left(\theta_a(t)-\omega_a t\right)
        \right]
    \right] .
    \label{eq:coherent-exact}
\end{align}

\begin{figure}[t]
    \centering
    \includegraphics[width=12cm]{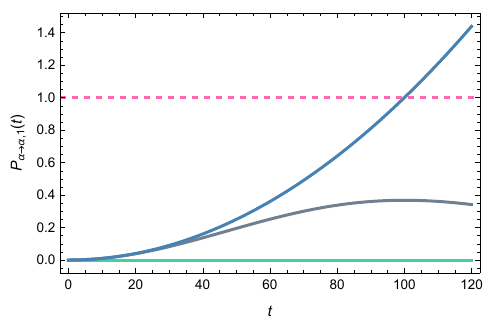}
    \caption{
        The conversion probability $P_{\alpha\to\alpha,1}(t)$ as a
        function of time. The blue curve shows the first-order
        perturbative result for an initial coherent state, while the gray
        curve shows the corresponding exact result. The green curve shows
        the single-graviton conversion probability
        $P_{g\to p}(t)=\sin^2 q(t)$. The magenta dashed line indicates the
        unitarity bound, $P_{\alpha\to\alpha,1}=1$. The parameters, chosen for illustrative purposes, are 
        $\omega_a=1.001\,\mathrm{Hz}$\,, $\omega_b=1.0\,\mathrm{Hz}$\,, $g=10^{-5}\,\mathrm{Hz}$\,, and
        $|\alpha|^2=10^6$. }
    \label{fig:coherent-comparison}
\end{figure}

Using the leading-order resonant solution $q(t)=gt$ and
$\theta_a(t)=\omega_a t$, Eq.~\eqref{eq:coherent-exact} reduces to
\begin{align}
    P_{\alpha\to\alpha,1}(t)
    =
    |\alpha|^2\sin^2(gt)
    \exp\left[
        -2|\alpha|^2\left(1-\cos gt\right)
    \right].
    \label{eq:coherent-leading}
\end{align}
At sufficiently early times, such that
\begin{align}
    |\alpha|^2(gt)^2\ll 1,
\end{align}
the exponential factor can be approximated by unity, yielding
\begin{align}
    P_{\alpha\to\alpha,1}(t)
    \simeq
    \left(|\alpha| g t \right)^2 \ .
    \label{eq:coherent-perturbative}
\end{align}
For our benchmark parameters, taking the coherent-state occupation number
to be $|\alpha|^2=10^{37}$ gives
\begin{align}
    P_{\alpha\to\alpha,1}(t)
    \simeq
    (30\,\mathrm{Hz}\times t)^2 \ .
    \label{eq:coherent-perturbative}
\end{align}
This reproduces the perturbative result obtained in~\cite{Tobar:2023ksi}. 
At $t=1\,\mathrm{s}$, this expression exceeds unity, indicating the breakdown
of the perturbative approximation. By contrast, the exact conversion
probability remains below $0.4$ for the same parameters.

The same perturbative enhancement has been obtained for
graviton-phonon conversion~\cite{Tobar:2023ksi},
axion-photon conversion~\cite{Ikeda:2025qac}, and
graviton-photon conversion~\cite{Ikeda:2025uae}.
The perturbative expression grows proportionally to $|\alpha|^2$ and can
therefore exceed unity for sufficiently large $|\alpha|$. This apparent
violation of the unitarity bound signals the breakdown of the perturbative
approximation. By contrast, the exact expression in
Eq.~\eqref{eq:coherent-exact} remains bounded by unity. As noted below Eq.~(2.5), the parameters used in the following figures
are chosen for illustrative purposes. Figure~\ref{fig:coherent-comparison} compares the first-order perturbative
result with the exact conversion probability. At early times, the coherent
state enhances the conversion probability through the factor $|\alpha|^2$.
As time increases, however, the perturbative result exceeds unity, whereas
the exact result remains below the unitarity bound.

\begin{figure}[t]
    \centering
    \includegraphics[width=12cm]{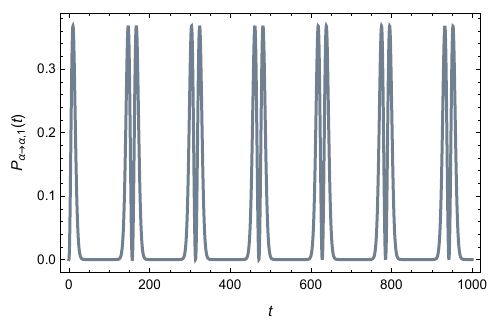}
    \caption{
The exact conversion probability
$P_{\alpha\to\alpha,1}(t)$ for a large coherent-state parameter.
The conversion occurs intermittently, appearing as narrow bursts separated
by intervals of strong suppression. In this regime, the first-order
perturbative approximation is no longer valid. The parameters are chosen to make the intermittent behavior clearly
visible and are $\omega_a=1.04\,\mathrm{Hz}$,
$\omega_b=1.0\,\mathrm{Hz}$, $g=10^{-5}\,\mathrm{Hz}$, and
$|\alpha|^2=10^8$.}
\label{fig:coherent-intermittency}
\end{figure}

For a larger coherent-state parameter, the perturbative approximation is no longer applicable. As shown in Fig.~\ref{fig:coherent-intermittency},
the exact conversion probability exhibits intermittent behavior, with narrow conversion bursts separated by intervals of strong suppression. This behavior originates from the phase-dependent interference contained
in the full quantum evolution and is not captured by first-order perturbation theory. Its observation could therefore provide a signature of the quantum dynamics of the graviton-phonon system.

The origin of the intermittency can be made explicit by writing
\begin{equation}
    P_{\alpha\to\alpha,1}(t)
    =
    |\alpha|^2\sin^2 q(t)\,
    \exp[-d^2(t)],
\end{equation}
where
\begin{equation}
    d^2(t)
    =
    \left|
        \alpha e^{i[\theta_a(t)-\omega_a t]}
        -\alpha\cos q(t)
    \right|^2
\end{equation}
is the squared distance between the two coherent states
entering the transition amplitude. For large $|\alpha|$, the overlap is
exponentially suppressed unless this distance becomes very small.
Consequently, the conversion probability is appreciable only during
short rephasing intervals, producing the observed intermittent bursts.

\subsection{Squeezed state}

We next consider the initial two-mode state
\begin{equation}
    \ket{\xi;0,0}
    \equiv
    \ket{\xi}_a\otimes\ket{0}_b,
\end{equation}
where the labels after the semicolon indicate the additional graviton and phonon excitations relative to the graviton squeezed vacuum. We define the normalized graviton-added squeezed state by
\begin{equation}
    \ket{\xi;1,0}
    \equiv
    \frac{1}{N}a^\dagger\ket{\xi;0,0},
    \qquad
    N=\cosh r.
\end{equation}
Under free evolution, the squeezing parameter evolves as
\begin{equation}
    \xi(t)=\xi e^{2i\omega_a t}.
\end{equation}

The transition amplitude to the final state $\ket{\xi(t);0,1}$ is therefore
\begin{align}
    \mathcal{A}_{\xi\to\xi,1}(t)
    &\equiv
    \bra{\xi(t);0,1}U(t)\ket{\xi;1,0}
    \nonumber\\
    &=
    \frac{1}{N}
    \bra{\xi(t);0,0}
    b\,U(t)\,a^\dagger
    \ket{\xi;0,0}
    \nonumber\\
    &=
    \frac{1}{N}
    \bra{\xi(t);0,0}
    b\,R_aR_bT\,a^\dagger
    \ket{\xi;0,0}.
\end{align}
The squeezed-vacuum state of the graviton mode can be expanded as
\begin{align}
    \ket{\xi}_a
    &=
    \frac{1}{\sqrt{\cosh r}}
    \sum_{n=0}^{\infty}
    \frac{\sqrt{(2n)!}}{2^n n!}
    \left(-e^{i\phi}\tanh r\right)^n
    \ket{2n}_a
    \nonumber\\
    &\equiv
    \sum_{n=0}^{\infty}
    c_n\ket{2n}_a .
\end{align}
where we used $\xi = r e^{-i\phi}$. Applying $Ta^\dagger$, we obtain
\begin{align}
    T a^\dagger |\xi\rangle_a
    =
    \sum_{n=0}^\infty
    c_n \sqrt{2n+1}\,T|2n+1\rangle_a .
\end{align}
Using
\begin{align}
    T a^\dagger
    =
    \left(
        \cos q\,a^\dagger
        +e^{-i\chi}\sin q\,b^\dagger
    \right)T,
\end{align}
we find
\begin{align}
    T a^\dagger |\xi\rangle_a
    =
    \sum_{n=0}^\infty
    c_n \sqrt{2n+1}\,
    \frac{1}{\sqrt{(2n+1)!}}
    \left(
        \cos q\,a^\dagger
        +e^{-i\chi}\sin q\,b^\dagger
    \right)^{2n+1}
    |0\rangle_a .
\end{align}
Since the bra state contains one phonon, only the terms proportional to $b^\dagger|2n\rangle$ contribute. Moreover,
\begin{align}
    \bra{\xi(t);0,0} b\,U\,a^\dagger \ket{\xi;0,0}
    =
    e^{-i\theta_b}
    \bra{\xi(t)e^{-2i\theta_a};0,0}
    b\,T a^\dagger
    \ket{\xi;0,0} .
\end{align}
Hence, the relevant part of $Ta^\dagger\ket{\xi;0,0}$ is
\begin{align}
    T a^\dagger \ket{\xi;0,0}
    =
    e^{-i\chi}\sin q
    \sum_{n=0}^\infty
    c_n (2n+1)(\cos q)^{2n}
    b^\dagger \ket{2n,0} .
\end{align}
Therefore, the amplitude becomes
\begin{align}
    \bra{\xi(t);0,0} b\,U\,a^\dagger \ket{\xi;0,0}
    =
    \frac{
        e^{-i(\chi+\theta_b)}\sin q
    }{
        \cosh r
        \left(
            1-\tanh^2 r\,\cos^2 q\,e^{2i\omega_a t-2i\theta_a}
        \right)^{3/2}
    } \,.
\end{align}
Here, we used
\begin{align}
    \sum_{n=0}^\infty
    (2n+1)\frac{(2n)!}{2^{2n}(n!)^2}x^n
    =
    \frac{1}{(1-x)^{3/2}} \,.
\end{align}
Finally, the conversion probability is
\begin{align}
    P_{\xi;1,0\to\xi;0,1}(t)&\equiv
    |\bra{\xi(t);0,1}
        U
        \ket{\xi;1,0}|^2\nonumber\\
    &=
    \frac{
        \sin^2 q
    }{
        \cosh^4 r
        \left[
            1+\tanh^4 r\,\cos^4 q
            -2\tanh^2 r\,\cos^2 q
            \cos\left(2\theta_a-2\omega_a t\right)
        \right]^3
    } \,.
\end{align}
At lowest order in perturbation theory, this reduces to
\begin{align}
    P_{\xi;1,0\to\xi;0,1}(t)=
    \cosh^2 r\,\sin^2(gt)\,.
\end{align}
This result was first derived in
Refs.~\cite{Ikeda:2025qac,Ikeda:2025uae}.

\begin{figure}[t]
        \centering
        \includegraphics[width=12cm]{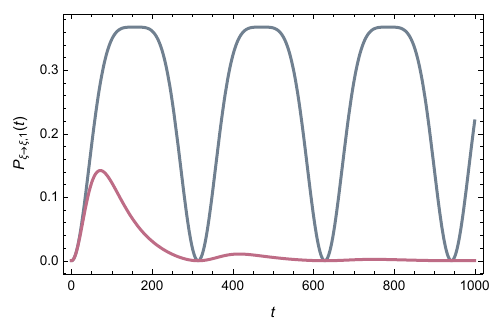}
     \vspace{-6pt}
        \caption{
Time evolution of the conversion probabilities for the coherent and
squeezed initial states. The gray curve shows
$P_{\alpha\to\alpha,1}(t)$ for the coherent state, while the red curve
shows $P_{\xi;1,0\to\xi;0,1}(t)$ for the squeezed state. The parameters
are chosen to highlight the difference between the coherent and squeezed
cases. Specifically, we set $|\alpha|^2=\cosh^2 r=10^6$, corresponding
to $r=7.6009$, so that the two cases have the same leading-order
perturbative enhancement. The remaining parameters are
$\omega_a=1.02\,\mathrm{Hz}$, $\omega_b=1.0\,\mathrm{Hz}$, and
$g=10^{-5}\,\mathrm{Hz}$.}

      \label{result}
        \vspace{0.7cm}
        \label{Fig3}
\end{figure}

Figure~\ref{Fig3} compares the time evolution of the conversion probabilities for coherent and squeezed initial states. We choose the parameters such that $|\alpha|^2=\cosh^2 r=10^6$, so that the two cases have the same leading-order perturbative enhancement. The departure from the perturbative behavior occurs earlier for the squeezed state than for the
coherent state. In addition, after the first peak, the squeezed-state conversion probability becomes strongly suppressed and remains close to zero at late times. This suppression is a non-perturbative feature of the
full quantum evolution.

Thus far, we have studied these non-perturbative effects within the rotating-wave approximation. The same formalism can, in principle, be applied to more general initial quantum states. Extending the analysis beyond the rotating-wave approximation is an interesting direction for future work.

\section{Conclusion}

We studied graviton-phonon conversion from a fully quantum perspective,
focusing on coherent and squeezed initial states. For an initial coherent
state, first-order perturbation theory predicts that the conversion
probability is proportional to $|\alpha|^2$, which corresponds to the mean
occupation number of the gravitational-wave mode. Because this occupation
number can be extremely large in the LIGO frequency band, around
$100\,\mathrm{Hz}$, the perturbative conversion probability can be strongly
enhanced. In this regime, however, the validity of the perturbative
approximation must be examined carefully.

Within the rotating-wave approximation, we solved the quantum dynamics
exactly. We found that, for a large coherent-state parameter, the exact
conversion probability is suppressed relative to the perturbative result
and remains consistent with the unitarity bound. We also found that the
conversion becomes intermittent, with narrow bursts separated by intervals
of strong suppression. This behavior arises from the full 
quantum evolution and is not captured by first-order perturbation theory.
Its observation could therefore provide a signature of the quantum dynamics
of the graviton-phonon system.

We also examined the squeezed-state case, for which a perturbative
enhancement was previously found in
Refs.~\cite{Ikeda:2025qac,Ikeda:2025uae}. The exact probability again satisfies the unitarity bound. Moreover, the departure from perturbative behavior occurs
earlier for the squeezed state than for the coherent state, and the
squeezed-state conversion probability becomes strongly suppressed after
its initial peak.

Although our discussion has focused on graviton-phonon conversion, the
formalism is more general. It would be interesting to apply the present
analysis to the squeezed-state enhancement of axion-photon
conversion~\cite{Ikeda:2025qac} and graviton-photon
conversion~\cite{Ikeda:2025uae}. We have not considered a concrete
experimental realization in this work. These extensions and the
development of a realistic experimental setup are left for future study.


\section*{Acknowledgments}
S.\ K. was supported by the Japan Society for the Promotion of Science (JSPS) KAKENHI Grant Numbers JP24K21548. J.\ S. was in part supported by JSPS KAKENHI Grant Numbers JP23K22491, JP24K21548, JP25H02186.

\bibliography{graviton} 

@article{Arimoto:2021cwc,
    author = "Arimoto, Makoto and others",
    title = "{Gravitational wave physics and astronomy in the nascent era}",
    eprint = "2104.02445",
    archivePrefix = "arXiv",
    primaryClass = "gr-qc",
    doi = "10.1093/ptep/ptab042",
    journal = "PTEP",
    volume = "2023",
    number = "10",
    pages = "10A103",
    year = "2023"
}

@article{LIGOScientific:2016aoc,
    author = "Abbott, B. P. and others",
    collaboration = "LIGO Scientific, Virgo",
    title = "{Observation of Gravitational Waves from a Binary Black Hole Merger}",
    eprint = "1602.03837",
    archivePrefix = "arXiv",
    primaryClass = "gr-qc",
    reportNumber = "LIGO-P150914",
    doi = "10.1103/PhysRevLett.116.061102",
    journal = "Phys. Rev. Lett.",
    volume = "116",
    number = "6",
    pages = "061102",
    year = "2016"
}

@article{Ito:2019wcb,
    author = "Ito, Asuka and Ikeda, Tomonori and Miuchi, Kentaro and Soda, Jiro",
    title = "{Probing GHz gravitational waves with graviton{\textendash}magnon resonance}",
    eprint = "1903.04843",
    archivePrefix = "arXiv",
    primaryClass = "gr-qc",
    reportNumber = "KOBE-COSMO-19-01",
    doi = "10.1140/epjc/s10052-020-7735-y",
    journal = "Eur. Phys. J. C",
    volume = "80",
    number = "3",
    pages = "179",
    year = "2020"
}

@article{Ito:2020wxi,
    author = "Ito, Asuka and Soda, Jiro",
    title = "{A formalism for magnon gravitational wave detectors}",
    eprint = "2004.04646",
    archivePrefix = "arXiv",
    primaryClass = "gr-qc",
    reportNumber = "KOBE-COSMO-20-07",
    doi = "10.1140/epjc/s10052-020-8092-6",
    journal = "Eur. Phys. J. C",
    volume = "80",
    number = "6",
    pages = "545",
    year = "2020"
}

@article{Grishchuk:1989ss,
    author = "Grishchuk, L. P. and Sidorov, Yu. V.",
    title = "{On the Quantum State of Relic Gravitons}",
    doi = "10.1088/0264-9381/6/9/002",
    journal = "Class. Quant. Grav.",
    volume = "6",
    pages = "L161--L165",
    year = "1989"
}

@article{Grishchuk:1990bj,
    author = "Grishchuk, L. P. and Sidorov, Yu. V.",
    title = "{Squeezed quantum states of relic gravitons and primordial density fluctuations}",
    doi = "10.1103/PhysRevD.42.3413",
    journal = "Phys. Rev. D",
    volume = "42",
    pages = "3413--3421",
    year = "1990"
}

@article{Preparata:1988am,
    author = "Preparata, Giuliano",
    title = "{QUANTUM MECHANICS OF A GRAVITATIONAL ANTENNA}",
    reportNumber = "Print-88-0353 (MILAN), Print-88-0515 (MILAN)",
    doi = "10.1007/BF02748965",
    journal = "Nuovo Cim. B",
    volume = "101",
    pages = "625",
    year = "1988"
}

@article{Grishchuk:1992gn,
    author = "Grishchuk, L. P.",
    title = "{Quantum mechanics of a solid state bar gravitational antenna}",
    doi = "10.1103/PhysRevD.45.2601",
    journal = "Phys. Rev. D",
    volume = "45",
    pages = "2601--2608",
    year = "1992"
}

@article{Manikandan:2025ykr,
    author = "Manikandan, Sreenath K. and Wilczek, Frank",
    title = "{Testing the coherent-state description of radiation fields}",
    doi = "10.1103/PhysRevA.111.033705",
    journal = "Phys. Rev. A",
    volume = "111",
    number = "3",
    pages = "033705",
    year = "2025"
}

@article{Manikandan:2025qgv,
    author = "Manikandan, Sreenath K. and Wilczek, Frank",
    title = "{Complementary Probes of Gravitational Radiation States}",
    eprint = "2505.11422",
    archivePrefix = "arXiv",
    primaryClass = "gr-qc",
    month = "5",
    year = "2025"
}

@article{Manikandan:2025hlz,
    author = "Manikandan, Sreenath K. and Wilczek, Frank",
    title = "{Probing Quantum Structure in Gravitational Radiation}",
    eprint = "2505.11407",
    archivePrefix = "arXiv",
    primaryClass = "gr-qc",
    month = "5",
    year = "2025"
}

@article{Kanno:2025how,
    author = "Kanno, Sugumi and Soda, Jiro and Taniguchi, Akira",
    title = "{Coherent State Description of Gravitational Waves from Binary Black Holes}",
    eprint = "2508.17947",
    archivePrefix = "arXiv",
    primaryClass = "gr-qc",
    reportNumber = "YITP-25-129, KOBE-COSMO-25-16",
    doi = "10.1103/kv1t-j27m",
    journal = "Phys. Rev. Lett.",
    volume = "136",
    number = "6",
    pages = "061404",
    year = "2026"
}

@article{Dyson:2013hbl,
    author = "Dyson, Freeman",
    title = "{Is a graviton detectable?}",
    doi = "10.1142/S0217751X1330041X",
    journal = "Int. J. Mod. Phys. A",
    volume = "28",
    pages = "1330041",
    year = "2013"
}

@article{Parikh:2020nrd,
    author = "Parikh, Maulik and Wilczek, Frank and Zahariade, George",
    title = "{The Noise of Gravitons}",
    note="arXiv:2005.07211 [hep-th]",
eprint = "2005.07211",
    archivePrefix = "arXiv",
    primaryClass = "hep-th",
    doi = "10.1142/S0218271820420018",
    journal = "Int. J. Mod. Phys. D",
    volume = "29",
    number = "14",
    pages = "2042001",
    year = "2020"
}

@article{Kanno:2020usf,
    author = "Kanno, Sugumi and Soda, Jiro and Tokuda, Junsei",
    title = "{Noise and decoherence induced by gravitons}",
    note="arXiv:2007.09838 [hep-th]",
eprint = "2007.09838",
    archivePrefix = "arXiv",
    primaryClass = "hep-th",
    reportNumber = "OU-HET-1065, KOBE-COSMO-20-12",
    doi = "10.1103/PhysRevD.103.044017",
    journal = "Phys. Rev. D",
    volume = "103",
    number = "4",
    pages = "044017",
    year = "2021"
}

@article{Parikh:2020kfh,
    author = "Parikh, Maulik and Wilczek, Frank and Zahariade, George",
    title = "{Quantum Mechanics of Gravitational Waves}",
    note="arXiv:2010.08205 [hep-th]",
eprint = "2010.08205",
    archivePrefix = "arXiv",
    primaryClass = "hep-th",
    doi = "10.1103/PhysRevLett.127.081602",
    journal = "Phys. Rev. Lett.",
    volume = "127",
    number = "8",
    pages = "081602",
    year = "2021"
}

@article{Parikh:2020fhy,
    author = "Parikh, Maulik and Wilczek, Frank and Zahariade, George",
    title = "{Signatures of the quantization of gravity at gravitational wave detectors}",
    note="arXiv:2010.08208 [hep-th]",
eprint = "2010.08208",
    archivePrefix = "arXiv",
    primaryClass = "hep-th",
    doi = "10.1103/PhysRevD.104.046021",
    journal = "Phys. Rev. D",
    volume = "104",
    number = "4",
    pages = "046021",
    year = "2021"
}

@article{Kanno:2021gpt,
    author = "Kanno, Sugumi and Soda, Jiro and Tokuda, Junsei",
    title = "{Indirect detection of gravitons through quantum entanglement}",
    note="arXiv:2103.17053 [gr-qc]",
eprint = "2103.17053",
    archivePrefix = "arXiv",
    primaryClass = "gr-qc",
    reportNumber = "KOBE-COSMO-21-06",
    doi = "10.1103/PhysRevD.104.083516",
    journal = "Phys. Rev. D",
    volume = "104",
    number = "8",
    pages = "083516",
    year = "2021"
}

@article{Arani:2026jyz,
    author = "Arani, Fateme Shojaei and Lamine, Brahim and Soda, Jiro",
    title = "{Cavity-QED Transducer of Gravitons}",
    eprint = "2603.27687",
    archivePrefix = "arXiv",
    primaryClass = "quant-ph",
    month = "3",
    year = "2026"
}

@article{Ikeda:2025uae,
    author = "Ikeda, Taiki and Kaku, Youka and Kanno, Sugumi and Soda, Jiro",
    title = "{Toward graviton detection via photon-graviton quantum state conversion}",
    eprint = "2507.01609",
    archivePrefix = "arXiv",
    primaryClass = "quant-ph",
    reportNumber = "KOBE-COSMO-25-12",
    doi = "10.1103/kxs5-yp85",
    journal = "Phys. Rev. D",
    volume = "112",
    number = "10",
    pages = "103504",
    year = "2025"
}

@article{Ikeda:2025qac,
    author = "Ikeda, Taiki and Kanno, Sugumi and Soda, Jiro",
    title = "{Enhancing photon-axion conversion probability with squeezed coherent states}",
    eprint = "2506.14354",
    archivePrefix = "arXiv",
    primaryClass = "quant-ph",
    reportNumber = "YITP-25-105, KOBE-COSMO-25-11",
    doi = "10.1103/c741-bwvs",
    journal = "Phys. Rev. D",
    volume = "113",
    number = "2",
    pages = "023539",
    year = "2026"
}

@article{Manikandan:2025dea,
    author = "Manikandan, Sreenath K. and Wilczek, Frank",
    title = "{Squeezed Quasinormal Modes from Nonlinear Gravitational Effects}",
    eprint = "2508.03380",
    archivePrefix = "arXiv",
    primaryClass = "gr-qc",
    month = "8",
    year = "2025"
}

@article{Guerreiro:2025sge,
    author = "Guerreiro, Thiago",
    title = "{Entanglement and squeezing of gravitational waves}",
    eprint = "2501.17043",
    archivePrefix = "arXiv",
    primaryClass = "gr-qc",
    doi = "10.1103/fn5d-mrsj",
    journal = "Phys. Rev. D",
    volume = "112",
    number = "10",
    pages = "L101904",
    year = "2025"
}

@article{Dorlis:2025zzz,
    author = "Dorlis, Panagiotis and Mavromatos, Nick E. and Sarkar, Sarben and Vlachos, Sotirios-Neilos",
    title = "{Superradiant Axionic Black-Hole Clouds as Seeds for Graviton Squeezing}",
    eprint = "2507.01689",
    archivePrefix = "arXiv",
    primaryClass = "gr-qc",
    reportNumber = "KCL-PH-TH/2025-15",
    month = "7",
    year = "2025"
}

@article{Dorlis:2025amf,
    author = "Dorlis, Pangiotis and Mavromatos, Nick E. and Sarkar, Sarben and Vlachos, Sotirios-Neilos",
    title = "{Squeezed gravitons from superradiant axion fields around rotating black holes}",
    eprint = "2507.23475",
    archivePrefix = "arXiv",
    primaryClass = "gr-qc",
    reportNumber = "KCL-PH-TH/2025-26",
    month = "7",
    year = "2025"
}

@article{Kanno:2018cuk,
    author = "Kanno, Sugumi and Soda, Jiro",
    title = "{Detecting nonclassical primordial gravitational waves with Hanbury-Brown{\textendash}Twiss interferometry}",
    eprint = "1810.07604",
    archivePrefix = "arXiv",
    primaryClass = "hep-th",
    reportNumber = "OU-HET-980, KOBE-COSMO-18-09",
    doi = "10.1103/PhysRevD.99.084010",
    journal = "Phys. Rev. D",
    volume = "99",
    number = "8",
    pages = "084010",
    year = "2019"
}

@article{Tobar:2023ksi,
    author = "Tobar, Germain and Manikandan, Sreenath K. and Beitel, Thomas and Pikovski, Igor",
    title = "{Detecting single gravitons with quantum sensing}",
    eprint = "2308.15440",
    archivePrefix = "arXiv",
    primaryClass = "quant-ph",
    reportNumber = "NORDITA 2023-040",
    doi = "10.1038/s41467-024-51420-8",
    journal = "Nature Commun.",
    volume = "15",
    number = "1",
    pages = "7229",
    year = "2024"
}

@article{Toccacelo:2026hcz,
    author = "Toccacelo, Kristian and Beitel, Thomas and Andersen, Ulrik Lund and Pikovski, Igor",
    title = "{Quantum State Characterization of Gravitational Waves via Graviton Counting Statistics}",
    eprint = "2602.09125",
    archivePrefix = "arXiv",
    primaryClass = "quant-ph",
    month = "2",
    year = "2026"
}

@article{Carney:2023nzz,
    author = "Carney, Daniel and Domcke, Valerie and Rodd, Nicholas L.",
    title = "{Graviton detection and the quantization of gravity}",
    eprint = "2308.12988",
    archivePrefix = "arXiv",
    primaryClass = "hep-th",
    reportNumber = "CERN-TH-2023-155",
    doi = "10.1103/PhysRevD.109.044009",
    journal = "Phys. Rev. D",
    volume = "109",
    number = "4",
    pages = "044009",
    year = "2024"
}

@inproceedings{Carney:2024dsj,
    author = "Carney, Daniel",
    title = "{Comments on graviton detection}",
    eprint = "2408.00094",
    archivePrefix = "arXiv",
    primaryClass = "gr-qc",
    month = "7",
    year = "2024"
}

@article{Albrecht:1992kf,
    author = "Albrecht, Andreas and Ferreira, Pedro and Joyce, Michael and Prokopec, Tomislav",
    title = "{Inflation and squeezed quantum states}",
    eprint = "astro-ph/9303001",
    archivePrefix = "arXiv",
    reportNumber = "IMPERIAL-TP-92-93-21",
    doi = "10.1103/PhysRevD.50.4807",
    journal = "Phys. Rev. D",
    volume = "50",
    pages = "4807--4820",
    year = "1994"
}

@article{Kanno:2025fpz,
    author = "Kanno, Sugumi and Soda, Jiro and Taniguchi, Akira",
    title = "{Binary gravitational waves as probes of quantum graviton states}",
    eprint = "2510.23326",
    archivePrefix = "arXiv",
    primaryClass = "gr-qc",
    reportNumber = "YITP-25-169, KOBE-COSMO-25-17",
    doi = "10.1103/yp4j-v7lg",
    journal = "Phys. Rev. D",
    volume = "113",
    number = "12",
    pages = "123542",
    year = "2026"
}

@article{Kanno:2017dci,
    author = "Kanno, Sugumi and Soda, Jiro",
    title = "{Infinite violation of Bell inequalities in inflation}",
    eprint = "1705.06199",
    archivePrefix = "arXiv",
    primaryClass = "hep-th",
    reportNumber = "KOBE-COSMO-17-08, KOBE-COSMO-17-08",
    doi = "10.1103/PhysRevD.96.083501",
    journal = "Phys. Rev. D",
    volume = "96",
    number = "8",
    pages = "083501",
    year = "2017"
}

@article{Kanno:2016gas,
    author = "Kanno, Sugumi and Shock, Jonathan P. and Soda, Jiro",
    title = "{Quantum discord in de Sitter space}",
    eprint = "1608.02853",
    archivePrefix = "arXiv",
    primaryClass = "hep-th",
    reportNumber = "KOBE-COSMO-16-09, KOBE-COSMO-16-09",
    doi = "10.1103/PhysRevD.94.125014",
    journal = "Phys. Rev. D",
    volume = "94",
    number = "12",
    pages = "125014",
    year = "2016"
}

@article{Kanno:2014bma,
    author = "Kanno, Sugumi and Shock, Jonathan P. and Soda, Jiro",
    title = "{Entanglement negativity in the multiverse}",
    eprint = "1412.2838",
    archivePrefix = "arXiv",
    primaryClass = "hep-th",
    reportNumber = "KOBE-TH-14-12, QGASLAB-14-06",
    doi = "10.1088/1475-7516/2015/03/015",
    journal = "JCAP",
    volume = "03",
    pages = "015",
    year = "2015"
}

@article{Kanno:2014lma,
    author = "Kanno, Sugumi and Murugan, Jeff and Shock, Jonathan P. and Soda, Jiro",
    title = "{Entanglement entropy of $\alpha$-vacua in de Sitter space}",
    eprint = "1404.6815",
    archivePrefix = "arXiv",
    primaryClass = "hep-th",
    reportNumber = "KOBE-TH-14-04, QGASLAB-14-02",
    doi = "10.1007/JHEP07(2014)072",
    journal = "JHEP",
    volume = "07",
    pages = "072",
    year = "2014"
}

@article{Das:2025kyn,
    author = "Das, Arunima and Parikh, Maulik and Wilczek, Frank and Wutte, Raphaela",
    title = "{Squeezed States in Gravity}",
    eprint = "2512.20601",
    archivePrefix = "arXiv",
    primaryClass = "gr-qc",
    month = "12",
    year = "2025"
}

@article{Kanno:2019gqw,
    author = "Kanno, Sugumi",
    title = "{Nonclassical primordial gravitational waves from the initial entangled state}",
    eprint = "1905.06800",
    archivePrefix = "arXiv",
    primaryClass = "hep-th",
    reportNumber = "OU-HET-1017",
    doi = "10.1103/PhysRevD.100.123536",
    journal = "Phys. Rev. D",
    volume = "100",
    number = "12",
    pages = "123536",
    year = "2019"
}

@article{Schumaker:1986tlu,
    author = "Schumaker, Bonny L.",
    title = "{Quantum mechanical pure states with gaussian wave functions}",
    doi = "10.1016/0370-1573(86)90179-1",
    journal = "Phys. Rept.",
    volume = "135",
    number = "6",
    pages = "317--408",
    year = "1986"
}

@article{Maldacena:2015bha,
    author = "Maldacena, Juan",
    title = "{A model with cosmological Bell inequalities}",
    eprint = "1508.01082",
    archivePrefix = "arXiv",
    primaryClass = "hep-th",
    doi = "10.1002/prop.201500097",
    journal = "Fortsch. Phys.",
    volume = "64",
    pages = "10--23",
    year = "2016"
}
\bibliographystyle{unsrt}
\end{document}